\documentclass[aps,prb,twocolumn,groupedaddress,showpacs]{revtex4-1}
\usepackage{amsmath}
\usepackage{graphicx}
\usepackage{epstopdf}
\usepackage{bm}
\usepackage{amssymb}

\begin{document}

  \title{Phase transitions of the f\/ive-state clock model on the square lattice}
  \author{Y. Chen$^1$, Z. Y. Xie$^2$, and J. F. Yu$^{1,*}$\\}
  \affiliation{$^1$Department of Applied Physics, School of Physics and Electronics, Hunan University, Changsha 410082, China\\
   $^2$Department of Physics, Renmin University of China, Beijing 100872}

  %\date{\today}

  \begin{abstract}
    Using the tensor renormalization group method based on the higher-order singular value decomposition, we have studied the phase transitions of the five-state clock model on the square lattice. The temperature dependence of the specific heat indicates the system has two phase transitions, as verified clearly in the correlation function. By investigating the magnetic susceptibility, we can only obtained the upper critical temperature as $T_{c2}=0.9565(7)$. From the fixed-point tensor, we locate the transition temperatures at $T_{c1}=0.9029(1)$ and $T_{c2} = 0.9520(1)$, consistent with the MC and the DMRG results.
  \end{abstract}


  \maketitle

\section{Introduction}\label{introduction}

  The research on exotic phases and the phase transitions has always been one of the central topics in statistical and condensed matter physics.\cite{Minnhagen}  A typical example is the continuous $XY$ model on the square lattice, which has attracted much attention since Kosterlitz and Thouless (KT) \cite{KT1, KT2} discovered a phase transition at finite temperature without any symmetry breaking, obviously beyond Landau's symmetry breaking theory for continuous phase transitions. Its low temperature phase, characterized by the bound vortex-antivortex pairs, is called the quasi-ordered or topological phase, with a divergent correlation length.
  The transition to the high temperature disordered (or paramagnetic) phase is attributed to the unbinding of these topological pairs.\cite{KT1, KT2}

  Its discrete version, the $q$-state clock model exhibits even richer features in phase transitions, dependent on $q$, and arouses extensive interests and debates consequently.
  As well-known, when $q \leqslant 4$, the model has only one order-disorder transition of Ising type; if $q$ approaches infinity, it then transits to the continuous $XY$ model, with one KT transition.\cite{Elitzur}
   Furthermore, it is agreed that there exists a critical $q_c$, when $q \geqslant q_c$, the system has two transitions, separated by the topological KT phase in-between. Apparently, the discrete symmetry plays an important role on the transitions.
  There has been a large amount of studies on this model with different $q$'s.\cite{Savit, Ortiz, Einhorn, Cardy, Spencer,Tobochnik, Tomita, Rastelli,  Lapilli, Borisen1, Borisen2, Kim, Baek2, Jing}

  At present, a consensus is reached that $q_c$ equals $5$, while the nature of the phase transitions is still under debate. As for $q=5$, an early renormalization group (RG) analysis suggested two KT-type transitions.\cite{Kadanoff}
  Recent Monte Carlo (MC) simulations even gave different predictions among themselves.\cite{Kumano, Baek3, Borisen2, Tobochnik, Lapilli}
  Usually, the helicity modulus is investigated, for which shows a jump to zero as a characterization of the KT transition in the planar $XY$ model from the quasi-ordered to the disordered phase, while there are differences in the definition and precise calculation of the helical modulus.\cite{Kumano, Baek3, Lapilli}
  Meanwhile, the other main difficulty is to determine the transition temperatures precisely, owing to its topological nature inherent. Using the density matrix renormalization group (DMRG) method with size up to $L=256$, Chatelain\cite{Christophe} estimated the values  $T_{c1}=0.914(12)$ and $T_{c2}=0.945(17)$ for the $q=5$ case. By calculating the susceptibility, Borisenko et al.\cite{Borisen2} gave $T_{c1}=0.90514(9)$ and $T_{c2}=0.95147(9)$ by MC method with size up to $L=1024$. By investigating the helicity modulus, Kumano et al.\cite{Kumano} predicted $T_{c1}=0.908$ and $T_{c2} = 0.944$ in MC method with size up to $L=256$.

  In recent years, tensor network state method has developed rapidly and become one of the most powerful numerical tool to study phase transitions in both classical and quantum many-body systems.\cite{Levin, HCJiang, Gu, Xie1, HHZhao2, Xie2, HHZhao1}  And the tensor renormalization group method based on the higher-order singular value decomposition (abbreviated as HOTRG),\cite{Xie2}   has been applied to continuous $XY$ model,\cite{JFYu}  3-dimensional Ising model \cite{Xie2} and Pott model\cite{MPQin, WShun} successfully.

  In this article, we have investigated the phase transitions of the ferromagnetic five-state clock model on the square lattice by the HOTRG method. By inspecting the specific heat and spin-spin correlation, we confirmed that the five-state clock model indeed has two phase transitions, separated by a narrow interval characterized with a power law correlation, apparently different to low- and high-temperature Ising type phases. Then we applied a static weak external magnetic field, and calculated the magnetic susceptibility. The upper phase transition temperature is obtained, at $T_{c2}=0.9565(7)$. Finally, we located the two transition temperatures as $T_{c1}=0.9029(1)$ and $T_{c2} = 0.9520(1)$ from the fixed-point tensor. The results are consistent with  the MC \cite{Kumano,Borisen2} and the DMRG \cite{Christophe} predications.
    %At low temperature, another topological excitation, i.e. the domain walls may prevail over the vortices, which condensate while the vortices proliferate with increasing the temperature.\cite{Ortiz, Einhorn}
  %It is the complexity of this coexistence and competition, that causes the difficulties in determination of both its transition nature and the precise location.

  The article is organized as follows: Sec.~\ref{Model} introduces the two-dimensional $q$-state clock model and the numerical method we used. The results and discussions are presented in Sec.~\ref{Results}. At last, a brief summary is given in Sec.~\ref{Summary}.

\section{Model and Method}\label{Model}

  The Hamiltonian of the ferromagnetic $q$-state clock model with an external magnetic f\mbox{}ield is def\mbox{}ined as
  \begin{equation}
    H = - J \sum_{\langle ij \rangle}\cos(\theta_{i}-\theta_{j})-h\sum_{i}\cos\theta_{i}\,,
  \end{equation}
  where $ \langle ij \rangle $ denotes the summation over the nearest neighbors and $\theta_i$ is the angle of the spin at site $i$, which is one of the discrete set $\theta=\frac{2 \pi k}{q}$, $k = 0, 1, 2, \ldots, q - 1 $. $J$ is the coupling constant between neighboring spins, which is set to 1 for convenience. $h$ is the applied weak magnetic field in unit of $J/\mu$, where $\mu$ is the magnetic moment of each spin also set as 1.

  For classical statistical systems, the tensor network state method usually starts by expressing the partition function as a product of all local tensors
  \begin{equation}
    Z = \mathrm{Tr}(e^{-\beta H}) = \mathrm{Tr} \prod_i T_{l_i,r_i,u_i,d_i}\,,
  \end{equation}
  where indices $(l_i,r_i,u_i,d_i)$ denote the four legs (left, right, up, down) of the tensor at site $i$. And each local tensor $T$ is defined by
  \begin{equation}
    T_{l,r,u,d}=\sum_{\sigma} W_{\sigma,l} W_{\sigma,r} W_{\sigma,u} W_{\sigma,d}\,,
  \end{equation}
  where $W$'s come from the singular value decomposition of the Boltzmann factor,
  \begin{equation}
     e^{\beta[\cos (\theta_i-\theta_j) + h\cos(\theta_i)/4+h\cos(\theta_j)/4]} = WW^{\dag}.
  \end{equation}
  As sketched in Fig.~1 of Ref.~[\onlinecite{Xie2}], the HOTRG coarse-graining scheme is a real space renormalization, with each step reducing the system size by half.
  Iterating the process along $x$ and $y$ directions alternatively, we can finally obtain the partition function and the free energy as well.

  %Considering the translation invariance, one can also evaluate of the expectation value of the local Hamiltonian or other observables as discussed in Refs.~[\onlinecite{Xie2, HHZhao1}], such as the internal energy or the magnetization.
  %From which, one can further obtain the specific heat or the magnetic susceptibility, respectively.
  %Similarly, the spin-spin correlation can be dealt with properly, as shown below in Fig.~\ref{Correlation}.

  \begin{figure}[tbp]
    \begin{center}
      \includegraphics[width=8cm,clip,angle=0]{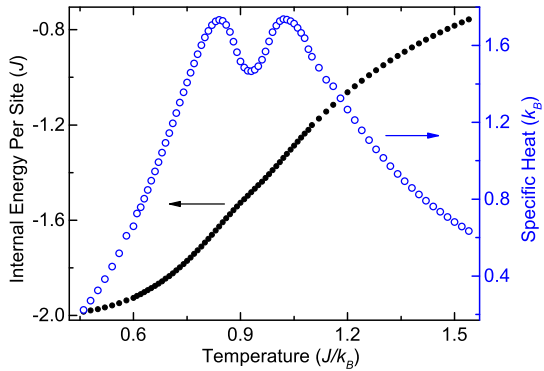}% Here is how to import EPS art [width=20mm,height=10mm][width=9cm,trim=0.5cm 21cm 0 2cm]
      \caption{\label{IE_SH}(Color online) Temperature dependence of the internal energy and the specific heat with $h=0$ and $D=40$.}
    \end{center}
  \end{figure}

  We should again note that the HOTRG is credited for handling big size systems. Only $\log_2 N$ steps are needed for the full contraction, given size $N$, which then could be very large to approximate the thermodynamic limit, and thus avoid the error inherent in the finite size scaling. In this work, we run the coarse-graining process until each physical quantity has converged. Normally, it takes 40 to 60 steps, so the system size is $2^{40}$ to $2^{60}$, approaching the infinity.
  One more thing should be mentioned is the size of the local tensor's legs, or the bond dimension denoted as $D$, is equal to $q$ initially. During each RG step, it is expanded quickly, and truncated to guarantee the process manageable. Usually, bigger the bond dimension, better the accuracy.

  \begin{figure}[tbp]
    \begin{center}
      \includegraphics[width=8cm,clip,angle=0]{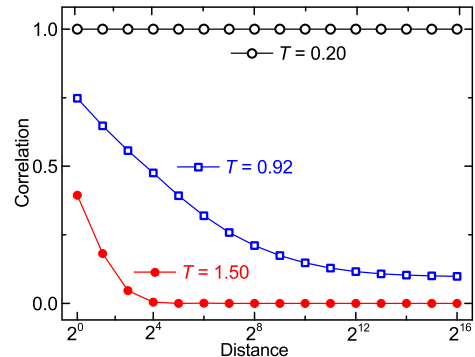}% Here is how to import EPS art [width=20mm,height=10mm][width=9cm,trim=0.5cm 21cm 0 2cm]
      \caption{\label{Correlation}(Color online)
      Spin-spin correlation along one direction. One representative temperature of each phase is shown. The results are obtained with $D=40$.}
    \end{center}
  \end{figure}

\section{Results and Discussions}\label{Results}

  As an illustration, we first show in Fig.~\ref{IE_SH} the internal energy and the specific heat without external magnetic field. In the specific heat, there are two obvious peaks around $0.8$ and $1.0$, indicating two possible transitions. To be more explicit, we calculate the correlation function at three representative temperatures. The correlation function is written as
  \begin{equation}
    C(r_{i}-r_{j})=\langle  \cos (\theta_{i})  \cos (\theta_{j})  \rangle + \langle  \sin (\theta_{i})  \sin (\theta_{j})  \rangle.
\end{equation}
  As sketched in Fig.~\ref{Correlation}, there are clearly three different behaviors, indicating two phase transitions separating the ordered and the disordered Ising type phases, as also suggested in Refs.~[\onlinecite{Elitzur, Ortiz, Einhorn, Kadanoff}].

  \begin{figure}[tbp]
    \begin{center}
      \includegraphics[width=8cm,clip,angle=0]{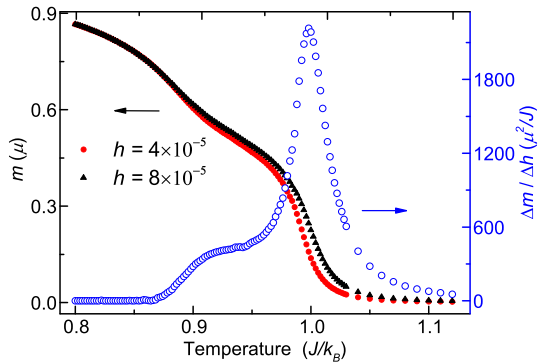}% Here is how to import EPS art [width=20mm,height=10mm][width=9cm,trim=0.5cm 21cm 0 2cm]
      \caption{\label{Mag_Sus}(Color online) Magnetization with existence of weak external magnetic fields, and the corresponding magnetic susceptibility as defined in Eq.~(\ref{suscept}) with $D=40$.}
    \end{center}
  \end{figure}

  The specific heat alone can not be used to determine the transition points, because it is derivatively continuous as in the $2D$ classical $XY$ model.
  We then turn to investigate the system's response to a static weak external magnetic field. We first evaluate the magnetization per spin, as shown in Fig.~\ref{Mag_Sus}. Because of the discrete orientations of the spin, the usually used magnetization exhibits significant fluctuations. Thus, we adopt the definition of the magnetization used in Refs.~[\onlinecite{Borisen1, Borisen2, Tomita, Rastelli, Tobochnik, Lapilli}] as %, Baek2}] as
  \begin{equation}
    m = \sqrt{ \langle \cos\theta \rangle ^2 + \langle \sin\theta \rangle ^2 } \,.
  \end{equation}
  Figure~\ref{Mag_Sus} shows the temperature dependence of the magnetization with two tentative magnetic fields. In the low temperature, the system is in the ordered phase with the magnetization approaches to $1$. In the high temperature, the system is in the disordered phase with the magnetization equals to $0$. In the critical phase, the magnetization takes values between $0$ and $1$. This behavior is similar to $XY$ case.\cite{JFYu}

  We adopt the magnetic susceptibility
  \begin{equation}\label{suscept}
    \chi=\frac{\Delta m}{\Delta h}=\frac{m(h_{2})-m(h_{1})}{h_{2}-h_{1}} \,,
  \end{equation}
  to determine the phase transition point, as plotted in Fig.~\ref{Mag_Sus}, one can observe a clear peak, signifying its analogy to the KT transition in the $XY$ model. The peak appears at $0.9990$ with the applied magnetic field at near $6\times10^{-5}$. The magnetic susceptibility decays to zero exponentially above the peak temperature, indicating a phase transition. However, the lower one does not manifest itself in the same way.

  \begin{figure}[tbp]
    \begin{center}
      \includegraphics[width=8cm,clip,angle=0]{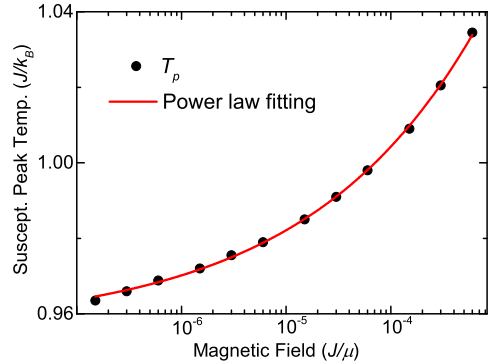}% Here is how to import EPS art [width=20mm,height=10mm][width=9cm,trim=0.5cm 21cm 0 2cm]
      \caption{\label{Tp_sus}(Color online) The peak position of the magnetic susceptibility with respect to the external field, and a power law fitting is also performed for extrapolation to zero-field limit. The critical point is obtained as  $T_{c2}=0.9561(10)$ for $D=40$ case.}
    \end{center}
  \end{figure}

  \begin{figure}[tbp]
    \begin{center}
      \includegraphics[width=8cm,clip,angle=0]{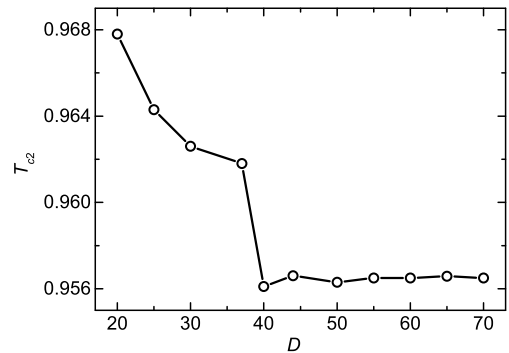}% Here is how to import EPS art [width=20mm,height=10mm][width=9cm,trim=0.5cm 21cm 0 2cm]
      \caption{\label{TCD} The upper phase transition temperature with respect to the bond dimension in the five-clock model by the magnetic susceptibility calculation.}
    \end{center}
  \end{figure}

  Following the procedure in Ref.~[\onlinecite{JFYu}], we can determine the transition temperature by locating the peak positions of the magnetic susceptibility with respect to the external fields, and then extrapolating to the zero-field limit. As only one peak shows up, the upper transition between the quasi-ordered and the disordered phases is then predicted at $T_{c2}=0.9561(10)$ by a power law fitting with $D=40$, as presented in Fig.~\ref{Tp_sus},
  \begin{equation}
    T_p-T_{c2}=ah^b\,,
  \end{equation}
   where $a=0.5666(310)$ and $b=0.2677(84)$. Here, we can see the behavior is similar to the KT transition in $XY$ model,\cite{JFYu} indicating it is also a KT-type transition. Further, we calculated the upper transition temperature with respect to different bond dimensions, as shown in Fig.~\ref{TCD}, the upper transition point gets converged at $T_{c2}=0.9565(7)$ with the bond dimension up to $70$.

For the lower transition point, we turn to focus on the local fixed-point tensor. At each coarse-graining step, an optimized isometric matrix would be obtained to truncate the local tensors. Eventually, each local tensor will flow to a corresponding fixed-point tensor. One can fetch information from the fixed-point tensors, as first introduced in  Ref.~[\onlinecite{Gu}] to identify different phases of Ising model, a gauge invariant quantity $X$ is defined as

\begin{equation}\label{XE}
    X=\frac{(\sum_{ru}T_{rruu})^{2}}{\sum_{lrud}T_{rlud}T_{lrdu}},
  \end{equation}
which $X$ can present the degeneracy of the system. It has also been used to locate the two  transition points of the six-state clock model.\cite{Jing}

  \begin{figure}[tbp]
    \begin{center}
      \includegraphics[width=8cm,clip,angle=0]{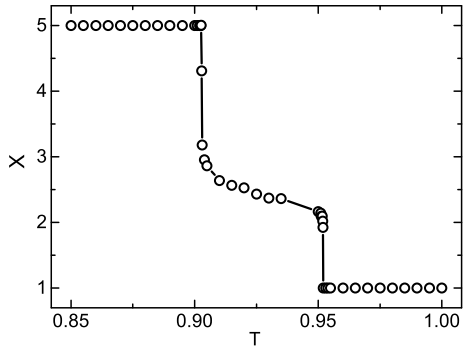}% Here is how to import EPS art [width=20mm,height=10mm][width=9cm,trim=0.5cm 21cm 0 2cm]
      \caption{\label{X} Temperature dependence of $X$ obtained from the fixed-point tensors. $X$ shows two jumps at the lower and upper phase transition, $T_{c1}=0.9029(1)$ and $T_{c2}=0.9520(1)$. The results are obtained with $D=75$ as an illustration.}
    \end{center}
  \end{figure}

\begin{figure}[tbp]
    \begin{center}
      \includegraphics[width=8cm,clip,angle=0]{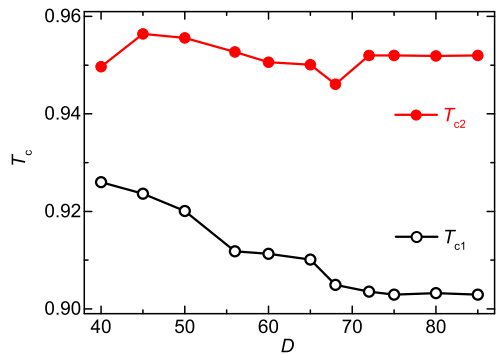}% Here is how to import EPS art [width=20mm,height=10mm][width=9cm,trim=0.5cm 21cm 0 2cm]
      \caption{\label{XTCD}(Color online) The lower and upper transition temperatures with respect to bond dimensions in the five-state clock model from $X$ calculation. }
    \end{center}
  \end{figure}

  As shown in Fig.~\ref{X}, $X$ equals $5$ in the ordered phase and $1$ in the disordered phase. In the critical phase, $X$ takes value between $5$ and $1$. We found two jumps occur at $0.9029(1)$ and $0.9520(1)$ corresponding to the lower and upper phase transition temperatures for $D=75$. Once again, as depicted in Fig.~\ref{XTCD}, we performed the bond dimension scaling to extrapolate the converged transition temperatures as $T_{c1}=0.9029(1)$ and $T_{c2}=0.9520(1)$ by calculating $X$ from the fixed-point tensor.
  The results agree well with the aforementioned MC \cite{Kumano, Borisen2} and the DMRG \cite{Christophe} results.

\section{Summary}\label{Summary}

  Briefly, we have studied the phase transitions of the ferromagnetic five-state clock model on the square lattice, using the HOTRG algorithm to calculate the thermodynamic observables with the existence of a weak external magnetic field and the fixed-point tensor.

  First, by calculating the specific heat and the correlation function, we confirmed that there are two phase transitions separating three phases: the low-temperature ordered phase, the high-temperature disordered phase, and the quasi-ordered phase in-between.
  Furthermore, by investigating the peak temperatures of the magnetic susceptibility with the weak external magnetic fields, the upper transition temperature could be determined. Through the upper transition temperatures with different $D$, we obtained the convergent upper transition temperature at $T_{c2}=0.9565(7)$. In order to locate the lower temperature, we extracted $X$ from the fixed-point tensors, which shows two jumps corresponding to two phase transitions of five-state clock model. We found the convergent transition temperatures, $T_{c1}=0.9029(1)$ and $T_{c2}=0.9520(1)$, which are consistent to the MC\cite{Kumano, Borisen2} and the DMRG\cite{Christophe} results.

\section{Acknowledgement}\label{Acknowledgement}

  This work was supported by the Fundamental Research Funds for the Central Universities (No. 531107040857) and the Natural Science Foundation of Hunan Province (No. 851204035). The authors are grateful to Profs. L. M. Tang and Q. Wan of Hunan University for computing resources.

\bibliography{clockmodeldraft}

\end{document}